\documentclass{aastex}
\usepackage{emulateapj5}
\usepackage{apjfonts}

\begin{document}

\title{Models of Type I X-ray Bursts from 4U~1820-30}

\author{Andrew Cumming} \affil{Hubble Fellow, UCO/Lick Observatory,
University of California, Santa Cruz, CA 95064; cumming@ucolick.org}

\begin{abstract}
I present ignition models for Type I X-ray bursts and superbursts from
the ultracompact binary 4U~1820-30. A pure helium secondary is usually
assumed for this system (which has an orbital period $\approx 11\ {\rm
min}$), however some evolutionary models predict a small amount of
hydrogen in the accreted material (mass fraction $X\sim 0.1$). I show
that the presence of hydrogen significantly affects the Type I burst
recurrence time if $X\gtrsim 0.03$ and the CNO mass fraction $\gtrsim
3\times 10^{-3}$. When regularly bursting, the predicted burst
properties agree well with observations. The observed $2$--$4$ hour
recurrence times are reproduced for a pure He companion if the
time-averaged accretion rate is $\dot{\left<M\right>}\approx
7$--$10\times 10^{-9}\ M_\odot\ {\rm yr^{-1}}$, or a hydrogen-poor
companion if $\dot{\left<M\right>}\approx 4$--$6\times 10^{-9}\
M_\odot\ {\rm yr^{-1}}$. This result provides a new constraint on
evolutionary models. The burst energetics are consistent with complete
burning, and spreading of the accreted fuel over the whole stellar
surface. Models with hydrogen predict $\sim 10$\% variations in burst
fluence with recurrence time, which perhaps could distinguish the
different evolutionary scenarios. I use the $\dot M$'s determined by
matching the Type I burst recurrence times to predict superburst
properties. The expected recurrence times are $\approx 1$--$2$ years
for pure He accretion (much less than found by Strohmayer and Brown),
and $\approx 5$--$10$ years if hydrogen is present. Determination of
the superburst recurrence time would strongly constrain the local
accretion rate and thermal structure of the neutron star.
\end{abstract}



\section{Introduction}

Type I X-ray bursts are well-understood as thermonuclear flashes on
the surface of an accreting neutron star (see Lewin, van Paradijs, \&
Taam 1995; Strohmayer \& Bildsten 2003 for reviews). The basic theory
was outlined more than twenty years ago (for an overview, see
Fujimoto, Hanawa, \& Miyaji 1981; Bildsten 1998), successfully
explaining the typical burst energies ($10^{39}$--$10^{40}\ {\rm
ergs}$), durations ($\sim 10$--$100\ {\rm s}$), and recurrence times
(hours to days). However, detailed comparisons of theory and
observations have had mixed success (Fujimoto et al.~1987; van
Paradijs, Penninx, \& Lewin 1988; Bildsten 2000; Galloway et
al.~2003).

The ultracompact binary 4U~1820-30 ($P_{\rm orb}=11.4\ {\rm min}$;
Stella, Priedhorsky, \& White 1987) is a promising system for such a
comparison. It has a known distance, being located in the metal-rich
globular cluster NGC~6624 ($[$Fe$/$H$]\approx -0.4$, distance $7.6\pm
0.4\ {\rm kpc}$); Rich, Minniti, \& Liebert 1993; Kuulkers et
al.~2003). It undergoes a regular $\approx 176$ day accretion cycle
(Priedhorsky \& Terrell 1984), switching between high and low states
differing by a factor $\approx 3$ in accretion rate (e.g.~see the {\it
RXTE}/ASM light curve in Figure 1 of Strohmayer \& Brown 2002). In the
low state, regular Type I bursts are seen ($\pm 23$ days around the
minimum luminosity; Chou \& Grindlay 2001), with recurrence times
$\approx 2$--$4$ hours (Clark et al.~1976; Clark et al.~1977; Haberl
et al.~1987; Cornelisse et al.~2003). No bursts are seen during the
rest of the cycle, implying a ``non-bursting'' mode of burning
(e.g.~Bildsten 1995). In the low state, 4U~1820-30 has also shown an
extremely energetic ($\sim 10^{42}\ {\rm erg}$) ``superburst'', likely
due to deep ignition of a carbon layer (Strohmayer \& Brown 2002; see
Kuulkers et al.~2002 and Strohmayer \& Bildsten 2003 for a summary of
superburst properties).

Different evolutionary scenarios for 4U~1820-30 have been
proposed. Those involving direct collision of a red giant and a
neutron star (Verbunt 1987), or formation of a neutron star-main
sequence binary by tidal capture (Bailyn \& Grindlay 1987) or an
exchange interaction (Rasio, Pfahl, \& Rappaport 2000) followed by a
common envelope phase, lead to accretion of pure helium from the
helium core of the red giant. A different picture is that the mass
transfer starts just after central hydrogen exhaustion (Tutukov et
al.~1987; Fedorova \& Ergma 1989; Podsiadlowski, Rappaport, \& Pfahl
2002), in which case the accreted material is again mostly helium, but
contains some hydrogen ($\approx 5$--$35$\% by mass; Fedorova \& Ergma
1989; Podsiadlowski, Rappaport, \& Pfahl 2002).

The observed energetics and recurrence times of the Type I bursts from
4U~1820-30 fit well with a picture of accumulation and burning of
helium-rich material. The ratio of persistent fluence (persistent flux
integrated over the burst recurrence time) to burst fluence is
$\alpha\approx 120$ (Haberl et al.~1987). For a gravitational energy
$GM/R\approx 200$ MeV per nucleon, this implies a nuclear energy
release $\approx 1.6$ MeV per nucleon, as expected for helium burning
to iron group elements. The mass of helium required to power the
observed energetics is roughly the mass accreted in the burst
recurrence time at the inferred accretion rate. The burst fluence
$\approx 3.5\times 10^{-7}\ {\rm erg\ cm^{-2}}$ (Haberl et al.~1987)
implies a total burst energy $2.5\times 10^{39}\ {\rm erg}\ (d/7.6\
{\rm kpc})^2$, or a mass of helium $\Delta M\approx 1.6\times 10^{21}\
{\rm g}$. The persistent luminosity when bursts are seen is
$L_X\approx 2.8\times 10^{37}\ {\rm erg\ s^{-1}}\ (d/7.6\ {\rm
kpc})^2(F_X/4\times 10^{-9}\ {\rm erg\ cm^{-2}\ s^{-1}})$, giving an
accretion rate $\dot M\approx 1.5\times 10^{17}\ {\rm g\
s^{-1}}\approx 2.4\times 10^{-9}\ M_\odot\ {\rm yr^{-1}}$ for a $1.4\
M_\odot$ neutron star with radius $R=10\ {\rm km}$. At this rate, a
mass $\Delta M$ is accreted in 3 hours, in excellent agreement with
the observed recurrence times.

A previous comparison of theory with observations of 4U~1820-30 was
carried out by Bildsten (1995, hereafter B95), who conducted
time-dependent simulations of pure helium burning on accreting neutron
stars. He found good agreement with observed energetics and recurrence
times, but for a somewhat hotter base temperature than expected. In
this paper, I calculate ignition models for Type I bursts from
4U~1820-30, and survey the conditions necessary to ignite helium at
the inferred mass $\Delta M$. I allow for a small amount of hydrogen
in the accreted material, and show that this has a large effect on the
burst recurrence time, because of extra heating from hot CNO hydrogen
burning as the layer accumulates. Finally, I discuss what we might
learn from simultaneous modelling of Type I bursts and
superbursts. The plan of the paper is as follows. The main results are
presented in \S 2. I calculate the critical mass needed for ignition,
and show the effect of hydrogen on recurrence times and energetics. I
compare these results to observations of Type I bursts, and to
evolutionary models. Finally, I use the constraints from Type I bursts
to construct ignition models for superbursts from 4U~1820-30, and
compare to the model of SB02. I summarize in \S 3, and discuss the
remaining open issues.

\section{Ignition Models and Comparison to Observations}

\subsection{Calculation of Ignition Conditions}

The physics and different regimes of nuclear burning on accreting
neutron stars have been described by several authors (for reviews, see
Lewin, van Paradijs, \& Taam 1995; Bildsten 1998). Here, I follow the
X-ray burst ignition calculations of Cumming \& Bildsten (2000,
hereafter CB00). I adopt plane-parallel coordinates in the thin
accreted layer, and work in terms of the column depth $y=P/g$, where
$dy=-\rho dz$, $P$ is the hydrostatic pressure, $\rho$ is the density,
and $g$ the surface gravity. I assume a $1.4\ M_\odot$ neutron star
with radius $R=10\ {\rm km}$, giving $g=(1+z)GM/R^2=2.45\times
10^{14}\ {\rm cm\ s^{-2}}$ where the gravitational redshift is
$z=1-(1-2GM/Rc^2)^{-1/2}=0.31$ (similar to the value $\approx 0.35$
recently inferred from absorption lines in {\it XMM} burst spectra of
EXO~0748-676; Cottam, Paerels, \& Mendez 2002). I include general
relativistic effects when calculating observable quantities. Cumming
et al.~(2002) showed that the Newtonian equations of CB00 have the
same form under general relativity, as long as $y$ and $\dot m$ refer
to rest mass rather than gravitational mass.

The thermal profile of the accumulating layer is described by the heat
equation $F=(4acT^3/3\kappa)(dT/dy)$, where the opacity $\kappa$ is
calculated as described by Schatz et al.~(1999), and the entropy
equation $dF/dy=-\epsilon$, where $F$ is the heat flux, and $\epsilon$
contains contributions from hot CNO hydrogen burning and compressional
heating. Compressional heating always plays a minor role at these
accretion rates, giving $\sim c_PT\approx 5k_BT/2\mu m_p\approx
0.02\,T_8\ {\rm MeV}$ per nucleon. The hot CNO energy production rate
is $\epsilon_H=5.8\times 10^{13}\ {\rm erg\ g^{-1}\ s^{-1}}\ (Z/0.01)$
(Hoyle \& Fowler 1965), where $Z$ is the CNO mass fraction. I follow
the changing hydrogen abundance with depth as described in CB00 (see
eqn.~[3] of that paper). Unless otherwise stated, I set
$Z=0.01$. Triple alpha burning during accumulation is not included,
this has only a small effect on the ignition conditions. I assume the
individual CNO abundances are those of the equilibrium hot CNO cycle
($^{14}$O and $^{15}$O). In the pure He case, the accreted $^{14}$N
survives to a density $\approx 10^6\ {\rm g\ cm^{-3}}$, when it
converts to $^{14}$C by electron capture (Hashimoto et al.~1986). For
the conditions here, no further processing occurs until helium ignites
via triple alpha.

To find the thermal profile, I repeatedly integrate downwards through
the accreted layer, iterating until $F=F_{\rm crust}$ at the base,
where $F_{\rm crust}=Q_{\rm crust}\left<\dot m\right>$ is the heat
flux from deeper layers. Here, $Q_{\rm crust}$ is the energy per
nucleon released in the crust from pycnonuclear and electron capture
reactions which escapes from the surface. Brown (2000) showed that
$Q_{\rm crust}\approx 0.1$ MeV per nucleon $\approx 10^{17}\ {\rm erg\
g^{-1}}$ for $\dot m\gtrsim 10^4\ {\rm g\ cm^{-2}\ s^{-1}}$. I assume
$F_{\rm crust}$ is set by the time-averaged local accretion rate
$\left<\dot m\right>$, rather than the instantaneous rate $\dot m$,
because the thermal time in the crust is much longer than the $\approx
6$ month luminosity cycle. Throughout this paper, I assume the
instantaneous rate when bursts are seen is half the time-average rate,
$\dot m=\left<\dot m\right>/2$ (as inferred from the {\it RXTE}/ASM
lightcurve; e.g.~Figure 1 of SB02).

The ignition column depth $y_{\rm ign}$ is found by increasing the
depth of the layer until the condition for ignition,
\begin{equation}\label{eq:ign}
{d\epsilon_{3\alpha}\over dT}={d\epsilon_{\rm cool}\over dT}
\end{equation}
(Fujimoto, Hanawa, \& Miyaji 1981; Fushiki \& Lamb 1987b; Bildsten
1998), is met at the base. Here $\epsilon_{3\alpha}$ is the triple
alpha energy production rate (Fushiki \& Lamb 1987a), and
$\epsilon_{\rm cool}=acT^4/3\kappa y$ is a local approximation to the
radiative cooling. The ignition criterion (\ref{eq:ign}) says that
ignition occurs when the heating from triple alpha is more temperature
sensitive than the cooling, and is large enough to change the
temperature of the layer. Following CB00, we include a multiplicative
factor of 1.9 in $\epsilon_{3\alpha}$ to account for proton captures
on $^{12}$C when hydrogen is present.

The advantage of this ``semi-analytic'' approach is that it allows a
survey of parameter space, while still giving a good estimate of the
ignition depth. CB00 compared the ignition conditions predicted by
equation (\ref{eq:ign}) with time-dependent models in the
literature. For mixed H/He ignition, the agreement was to
$10$--$30$\%, and better than a factor of two for pure He
ignitions. In addition, I have compared results for pure helium with
those of B95. For a given $\dot m$, I adjust the flux from the base to
match the base temperature specified by B95. For the $\dot m$'s
relevant to this paper, I find that the ignition column depth from
equation (\ref{eq:ign}) underpredicts the ignition column by
$10$--$30$\%.

I estimate the burst energy by assuming complete burning of the
accreted layer, and taking a nuclear energy release $Q_{\rm
nuc}=1.6+4.0\left<X\right>$ MeV per nucleon, where $\left<X\right>$ is
the mass-weighted mean $X$ in the layer at ignition. This expression
for $Q_{\rm nuc}$ includes 35\% energy loss in neutrinos during
hydrogen burning (Fujimoto et al.~1987). This is appropriate for
rp-process hydrogen burning up to iron group, and may be an
overestimate for small $X_0$, in which case we underestimate the burst
energy by up to $\approx 10$\% for $\left<X\right>\sim 0.1$. Assuming
the layer of fuel covers the whole star, the total burst energy is
$E_{\rm burst}=4\pi R^2y_{\rm ign}Q_{\rm nuc}/(1+z)$, where the
$(1+z)$ factor accounts for gravitational redshift.

\subsection{Effect of Hydrogen}

Hydrogen burning gives $\approx 7$ MeV per nucleon, so that even a
small amount of hydrogen burning during accumulation dominates the
$\approx 0.1$ MeV per nucleon coming from the crust. For pure helium
accretion, $F_{\rm crust}=\left<\dot m\right>Q_{\rm crust}\approx
10^{21}\ {\rm erg\ cm^2\ s^{-1}}\ \left<\dot m\right>_4Q_{17}$ sets
the temperature of the accumulating layer. If hydrogen is present, the
flux from the hot CNO cycle is $F_H=\epsilon_Hy_H=5.8\times 10^{21}\
{\rm erg\ cm^2\ s^{-1}}\ y_{H,8}\ (Z/0.01)$. Here, $y_H$ is the column
density of the layer which is burning hydrogen: this is either the
column depth at which hydrogen runs out, or the depth at which helium
ignites, whichever occurs first. Whether $F_H>F_{\rm crust}$ depends
on $y_H$ since this measures the amount of hydrogen burning. The time
to burn all the hydrogen is $t_H=E_HX_0/\epsilon_H$, or
\begin{equation}\label{eq:tH}
t_H=2.9\ {\rm hrs}\ \left({0.01\over Z}\right)\
\left({X_0\over 0.1}\right),
\end{equation}
where $E_H=6.0\times 10^{18}\ {\rm erg\ g^{-1}}$ is the energy release
per gram from hydrogen burning (unlike CB00, we include the energy
release from neutrinos here as given by Wallace \& Woosley 1981), and
$X_0$ is the initial hydrogen abundance. If the burst recurrence time
$t_{\rm recur}<t_H$, then helium ignition occurs before hydrogen
exhaustion, giving $y_H=\dot mt_{\rm recur}$. Then requiring
$F_H>F_{\rm crust}$ gives
\begin{equation}\label{eq:condition1}
Z>3.5\times 10^{-3}\ Q_{17}\ \left({t_{\rm recur}\over 3\
{\rm h}}\right).
\end{equation}
If $t_{\rm recur}>t_H$, all of the hydrogen is burned before helium
ignites, giving $y_H=\dot mt_H=1.2\times 10^8\ {\rm g\ cm^{-2}}\
(X_0/0.1)(\dot m_4/1.2)(0.01/Z)$. Then
\begin{equation}\label{eq:condition2}
X_0>0.034\ Q_{17}\left({\left<\dot m\right>\over 2\dot m}\right)
\end{equation}
is required for $F_H>F_{\rm crust}$. 

Equations (\ref{eq:condition1}) and (\ref{eq:condition2}) show that if
hydrogen burning is to affect the temperature profile in the layer,
both the amount of hydrogen and the metallicity must be large
enough. Evolutionary models for 4U~1820-30 with hydrogen give
$X\gtrsim 0.1$ (e.g.~Podsiadlowski et al.~2002), satisfying equation
(\ref{eq:condition2}). In addition, NGC~6624 is metal-rich
($[$Fe$/$H$]\approx -0.4$; Rich, Minniti, \& Liebert 1993) so that the
accreted metallicity satisfies equation (\ref{eq:condition1})
(although spallation in the accretion shock could reduce the
metallicity in the accreted layer; Bildsten, Salpeter, \& Wasserman
1992).

\subsection{Comparison with Observations}

Several authors have presented observations of regular bursting from
4U~1820-30 (the first source to exhibit X-ray bursts; Grindlay et
al.~1976). Clark et al.~(1976) saw 10 bursts with SAS-3 with a
recurrence time of $4.4\pm 0.17\ {\rm h}$ (a $4$\% variation), and
mean fluence $\approx 2.9\times 10^{-7}\ {\rm erg\ cm^{-2}}$
($1$--$10\ {\rm keV}$). The persistent flux was $\approx 1/5$ of the
peak flux seen in the high state. Further SAS-3 observations of the
transition from low to high state found 22 bursts (Clark et
al.~1977). The recurrence time decreased from $3.4$--$2.2$ hours as
the flux increased by a factor of $\approx 5$. The mean burst fluence
was $\approx 4\times 10^{-7}\ {\rm erg\ cm^{-2}}$. The decay time of
the bursts decreased slightly as flux increased. Haberl et al.~(1987)
saw 7 bursts in a 20 hour {\it EXOSAT} observation with a recurrence
time of $3.2\pm 0.05$ hours, mean fluence $(3.48\pm 0.16)\times
10^{-7}\ {\rm erg\ cm^{-2}}$, and $\alpha=142\pm 10$. The persistent
flux was $\approx 4.3\times 10^{-9}\ {\rm erg\ cm^{-2}\ s^{-1}}$. Most
recently, BeppoSAX/WFC accumulated $\approx 7\ {\rm Ms}$ of exposure
of the Galactic Center region, observing 49 bursts from 4U~1820-30
(Cornelisse et al.~2003; Kuulkers et al.~2003). These bursts show a
transition from regular to irregular bursting as persistent flux
increases (Cornelisse et al.~2003).

The simplest model is to consider pure helium accretion at the rate
inferred from the X-ray luminosity in \S 1, and to take $Q_{\rm
crust}=0.1$ MeV per nucleon, as found by Brown (2000). As a useful
reference point, I refer to this accretion rate as $\dot m_X\equiv
1.2\times 10^4\ {\rm g\ cm^{-2}\ s^{-1}}$. Figure 1 shows the
temperature profile at ignition for this case (lower curve). The
recurrence time and burst energy are $\approx 5$ times larger than the
observed values. Ignition occurs at a column depth $\approx 7\times
10^8\ {\rm g\ cm^{-2}}$, giving a recurrence time of $16.5\ {\rm h}$,
and burst energy $1.1\times 10^{40}\ {\rm erg}$. A model with hydrogen
gives much better agreement. The upper curve in figure 1 shows the
same model, but with $X_0=0.1$. The extra heating from hot CNO burning
leads to earlier ignition, $y\approx 2\times 10^8\ {\rm g\ cm^{-2}}$,
giving $t_{\rm recur}=4.4\ {\rm h}$ and $E_{\rm burst}=3.1\times
10^{39}\ {\rm erg}$.

Unfortunately, this result can not be taken as evidence for the
presence of hydrogen. A larger flux from below will also heat the
layer and lead to earlier ignition. For pure helium accretion at $\dot
m=\dot m_X$, I find that $Q_{\rm crust}=0.4$ MeV per nucleon is
required to bring the recurrence time down to $\approx 3$ hours. This
is larger than expected from heating in the crust (Brown
2000). However, additional heating might be provided by a previous
X-ray burst. Most of the energy released in a burst is radiated from
the surface of the star; however, some energy is conducted inwards,
and emerges on a longer timescale. This effect has been emphasized by
Taam and coworkers (Taam 1980; Taam et al.~1993), who refer to it as
``thermal inertia''. To estimate the size of this effect, I assume a
fraction $f$ of the nuclear energy flows inwards, and is released at a
constant rate between bursts. The extra flux is then $fE_{\rm
burst}/4\pi R^2t_{\rm recur}$, giving a contribution to $Q_{\rm
crust}$ of $\approx 0.1$ MeV per nucleon $(f/0.1)(3\ {\rm h}/t_{\rm
recur})(E_{39}/3)(\dot m_X/\dot m)$. This is likely an overestimate,
since the energy release decreases over time. A better estimate
requires time-dependent simulations of a sequence of bursts. For now,
I incorporate this uncertainty into $Q_{\rm crust}$.

An additional uncertainty is the $L_X$--$\dot m$ relation, which
allows some freedom in setting $\dot m$. Pure helium models nicely
agree with observations if $\dot m\approx 1.5$--$2\ \dot m_X$ (for
$Q_{\rm crust}=0.1$--$0.2$ MeV per nucleon). This is shown in Figure
\ref{fig:recur}, in which $t_{\rm recur}$ and $E_{\rm burst}$ are
plotted against $\dot m$. For each choice of $\dot m$, the
time-averaged rate is assumed to be $\left<\dot m\right>=2\dot m$. The
vertical bars at $\dot m=\dot m_X$ show the observed range of
recurrence time and energy. As we found above, hydrogen models require
$\dot m\approx \dot m_X$ to give agreement.

Table \ref{tab:1} shows the same result for some specific
models. Here, I choose $X_0$, $Z$, and $Q_{\rm crust}$, and then vary
$\dot m$ to match the recurrence time seen by Haberl et al.~(1987),
$t_{\rm recur}=3.2$ hours. There a few points worth noting. First, the
ignition depth in the helium models is sensitive to $Q_{\rm crust}$,
since this sets the temperature of the accumulating layer. As $Q_{\rm
crust}$ increases, $y_{\rm ign}$ decreases. Second, hydrogen models
show two regimes of behavior. If $t_{\rm recur}<t_H$ (see
eq.~[\ref{eq:tH}]), hydrogen burning occurs throughout the
accumulating layer. Then $y_{\rm ign}$ depends on the metallicity $Z$
which determines the amount of hot CNO burning, but not on $Q_{\rm
crust}$ or $\dot m$. As $Z$ decreases, $y_{\rm ign}$ increases. An
extreme case is Model 7 in Table \ref{tab:1} ($Z=10^{-3}$), in which
case little heating occurs and the ignition conditions are similar to
pure helium. If $t_{\rm recur}>t_H$, hydrogen runs out and ignition
occurs in a pure helium layer. Then $y_{\rm ign}$ is sensitive to
$\dot m$, explaining the steep increases in $t_{\rm recur}$ and
$E_{\rm burst}$ for the H models in Figure \ref{fig:recur} (see CB00
for more discussion of these different regimes).

In general, the burst energy is ${\it overpredicted}$ in these models,
by factors of up to $\approx 70$\%. This discrepancy is easily
accounted for by anisotropic burst emission (e.g.~Lapidus \& Sunyaev
1985; Fujimoto 1988) (although there may be other explanations, see
discussion in \S 3.1). For the remainder of this section, I
renormalize the predicted fluence to agree with the {\it EXOSAT} mean
value (Haberl et al.~1987).

Figure \ref{fig:f4} shows the expected {\it variations} in burst
fluence and recurrence time as a function of persistent flux. I show
hydrogen models with metallicity $Z=0.01$ ($X=0.1$ and $X=0.2$), a low
metallicity model $Z=10^{-3}$ ($X=0.1$), and a helium model with
$Q_{\rm crust}=0.1$. I assume $Q_{\rm crust}$ is independent of
accretion rate, not including any variations with $\dot m$ because of
thermal inertia. In each case, I normalize the $F_X$--$\dot m$
relation to give $t_{\rm recur}=3.2\ {\rm h}$ for $F_X=4.3\times
10^{-9}\ {\rm erg\ cm^{-2}\ s^{-1}}$ (Haberl et al.~1987). To do this,
I write $F_X=L_X/4\pi d^2=\dot m Q_{\rm grav} (R/d)^2 \xi_p^{-1}$,
where $Q_{\rm grav}=c^2z/(1+z)$ is the gravitational energy release
($221$ MeV per nucleon for $z=0.31$), and the factor $\xi_p$ accounts
for the anisotropy of the X-ray emission, as well as additional
uncertainties such as luminosity emerging at other wavelengths. Given
the $\dot m$ for the model with $t_{\rm recur}=3.2\ {\rm h}$, I set
\begin{equation}
\xi_p^{-1}\left({R/d\over 10\ {\rm km} @7.6\ {\rm
kpc}}\right)^2={1.1\over \dot m_4}\left({F_{X,-9}\over
4.3}\right)\left({221\ {\rm MeV}\over Q_{\rm grav}}\right)
\end{equation}
which gives the correct normalization. The values of
$\xi_p^{-1}(R_6/d_{7.6})^2$ implied are $0.5$ $(0.8)$ for the pure He
($X_0=0.1$) model. Since the thermal time in the crust is much longer
than the $\approx 6$ month luminosity cycle, I keep $\left<\dot
m\right>$ fixed as $F_X$ varies, at the value appropriate for $t_{\rm
recur}=3.2\ {\rm h}$.

The expected variation of recurrence time with $\dot m$ agrees
qualitatively with the observed behavior in that recurrence time drops
with increasing $F_X$. Clark et al.~(1976) found $t_{\rm recur}=4.4$
when the flux was $\approx 1/5$ of its peak value. Estimating the peak
flux to be roughly $3$ times the EXOSAT value, or $\approx 1.2\times
10^{-8}\ {\rm erg\ cm^{-2}\ s^{-1}}$, gives $F_X=2.4\times 10^{-9}\
{\rm erg\ cm^{-2}\ s^{-1}}$ for these observations, less than the
EXOSAT value and consistent with a longer $t_{\rm recur}$. However,
the dependence of $t_{\rm recur}$ on $\dot m$ observed by Clark et
al.~(1977) is not as steep as $\propto 1/\dot m$ predicted in the
models. They report that $t_{\rm recur}$ decreased from $3.4$ hours to
$2.2$ hours as the persistent flux increased by a factor of 5. In the
models, only the $X=0.1$ model shows a significant deviation from a
$1/\dot m$ law, but in the opposite direction. For $t_{\rm
recur}<t_H=3$ hours, $t_{\rm recur}$ falls off {\it more steeply} than
$1/\dot m$, opposite to the observed trend.

In Figure \ref{fig:f5}, I plot fluence against $t_{\rm recur}$
directly. The fluence is renormalized to agree with the mean EXOSAT
value, highlighting the variation in fluence with $t_{\rm
recur}$. Hydrogen models show a change in fluence of $\approx 10$\% as
the recurrence time varies over the observed range, whereas the helium
and low metallicity models have roughly constant fluence. When
hydrogen is present, the variation arises because the composition of
the layer at ignition changes according to the amount of hydrogen
burning that occurs during accumulation. I also plot the observed
values in Figure \ref{fig:f5}, showing the range of recurrence times
for the Clark et al.~(1977) observations, measurements for individual
bursts from Haberl et al.~(1987), and the mean fluence reported by
Clark et al.~(1976). There is a $\approx 30$\% variation in the
reported fluences. However, the error in the SAS-3 fluence
measurements is likely larger than this. For example, a reanalysis by
Clark (reported by Vacca et al.~1986) gave a $\approx 40$\% reduction
in the peak luminosities of individual bursts. Therefore any
conclusions drawn from Figure \ref{fig:f5} are probably
unwarranted. The comparison is complicated by intrinsic scatter in the
burst fluences. The scatter in the Haberl et al.~(1987) points in
Figure \ref{fig:f5} is $4$\%, comparable to the predicted variation
with recurrence time if hydrogen is present. Nonetheless, looking for
variations in fluence with recurrence time or persistent flux is a
promising way to distinguish between models with and without hydrogen.
It perhaps could be addressed with the BeppoSAX data (Cornelisse et
al.~2003).

\subsection{Comparison with Evolutionary Models}

The requirement that $\dot m\approx\dot m_X$ if hydrogen is present,
or $\dot m\approx 1.5$--$2\dot m_X$ for pure helium, is a new
constraint on evolutionary models. Evolutionary models for 4U~1820-30
predict values of $\dot M$ and $X_0$ (Fedorova \& Ergma 1989;
Podsiadlowski et al.~2002). Table \ref{tab:2} gives the resulting
ignition conditions for Type I bursts. There is a remarkable agreement
between the $\dot m$'s required to match Type I burst properties, and
those predicted by evolutionary models (although it should be noted
that these predicted $\dot m$'s are rather uncertain, e.g.~see
Podsiadlowski et al.~2002). For example, if the companion is a cold
helium white dwarf, then gravitational radiation angular momentum
losses give $\dot M_{GR}\approx 8.8\times 10^{-9}\ M_\odot\ {\rm
yr^{-1}}$ (for a companion mass of $0.068\ M_\odot$; Podsiadlowski et
al.~2000), or $\dot m\approx \left<\dot m\right>/2\approx 2.2\times
10^4\ {\rm g\ cm^{-2}\ s^{-1}}\ (R/10\ {\rm km})^{-2}$, consistent
with the values we find in Table 1 for $X=0$. In addition,
evolutionary models with hydrogen tend to have lower accretion rates
than pure helium models, by about a factor of two, so that in general
both are consistent with the data. There are particular models,
however, that do not give good agreement. For example, Podsiadlowski
et al.~(2002) give a hydrogen model with $\dot M=1.1\times 10^{-8}\
M_\odot\ {\rm yr^{-1}}$, which has a recurrence time of $1.4$ hours,
outside the observed range.

\subsection{Superburst Ignition Conditions}

A superburst was seen from 4U~1820-30 on 1999 Sep 9 (Strohmayer \&
Brown 2002). The source was in the low state, with $F_X=3.5\times
10^{-9}\ {\rm erg\ cm^{-2}\ s^{-1}}$. The total fluence of the
superburst was $2.7\times 10^{-4}\ {\rm erg\ cm^{-2}}$, giving a total
energy release of $\approx 2\times 10^{42}\ {\rm ergs}$ (almost a
factor of 1000 larger than a normal Type I burst). SB02 modelled the
superburst from 4U~1820-30, proposing that carbon production during
stable burning in the high state leads to a thick carbon/iron layer
(mass $\approx 10^{26}\ {\rm g}$, carbon mass fraction $X_C\gtrsim
0.3$) which ignites to give the superburst.

SB02 estimated the accretion rate for their model from the X-ray
luminosity, finding $\dot m\approx \dot m_X$. I now calculate ignition
conditions for the 4U~1820-30 superburst using the $\dot m$'s inferred
from Type I bursts. I follow the carbon ignition calculations of
Cumming and Bildsten (2001, herafter CB01), except I take the heavy
element to be $^{56}$Fe (CB01 considered much heavier elements than
iron, made during rp process burning of hydrogen). I integrate
downwards through the carbon/iron layer, assuming a constant heat flux
from the crust $F_{\rm crust}=Q_{\rm crust}\left<\dot m\right>$ (I
assume that the heat flux is the same as that heating the helium layer
between normal Type I bursts). I calculate the carbon ignition depth
using a similar criterion to equation (\ref{eq:ign}). To estimate the
superburst energy, I assume that carbon burning gives 1 MeV per
nucleon. The amount of carbon is uncertain, and depends on the
fraction of stable H/He burning, and the carbon yield. To allow a
direct comparison with SB02, I assume that the carbon mass fraction
$X_C$ is 30\% in these models.

Table 3 gives the resulting ignition conditions. The model numbers
correspond to those in Table 1, except for the model labelled
``SB02'', which has $\dot m=\dot m_X$. This gives $t_{\rm recur}=14$
years, and total energy release $\approx 4\times 10^{43}\ {\rm ergs}$,
similar to the values found by SB02. For the helium models, the larger
$\dot m$ than assumed by SB02 gives an order of magnitude shorter
recurrence time, $\approx 1$--$2$ years rather than $14\ {\rm years}$
(this sensitivity of recurrence time to $\dot m$ was noted by SB02,
see also Figure 3 of CB01). For the $X=0.1$ models, the recurrence
time is longer, $4$--$10$ years, depending on $Q_{\rm crust}$. This
shows that the superburst recurrence time, currently not well
constrained, is a powerful discriminant between different models.

Note that the total nuclear energy release can be much greater than
the observed superburst energy if neutrino emission plays an important
role. This was emphasized by SB02 for their model of the 4U 1820-30
superburst, in which only $\approx 10^{42}\ {\rm erg}$ (from a total
of $4\times 10^{43}\ {\rm ergs}$) is radiated from the
surface. However, the amount of neutrino emission is strongly
dependent on the peak temperature, and therefore carbon fraction and
ignition depth (CB01). Time-dependent models are therefore necessary
to predict the superburst fluence.

CB01 showed that for small carbon mass fractions, the carbon burns
stably as the layer accumulates. Table 3 gives the minimum $X_C$
required for unstable ignition for each model. The range of values is
$X_{C,{\rm min}}=0.08$--$0.2$. Less carbon is required for smaller
$Q_{\rm crust}$, as shown by CB01. The values of $X_{C,{\rm min}}$
given here are a little larger than those shown in CB01's Figure 2
because the less massive heavy element ($^{56}$Fe rather than
$^{104}$Ru) results in a shallower temperature gradient. The
recurrence time for $X_C=X_{C,{\rm min}}$ is $\approx 10$\% lower than
the recurrence time for $X_C=0.3$.

\section{Summary and Discussion}

\subsection{Type I bursts}

In general, the models of Type I X-ray bursts presented in \S 2 give
good agreement with observed recurrence times and energetics. The
observed $2$--$4$h recurrence times are reproduced for a pure He
companion if the time-averaged accretion rate is
$\dot{\left<M\right>}\approx 7$--$10\times 10^{-9}\ M_\odot\ {\rm
yr^{-1}}\ (R/10\ {\rm km})^2$, or a hydrogen-poor companion if
$\dot{\left<M\right>}\approx 4$--$6\times 10^{-9}\ M_\odot\ {\rm
yr^{-1}}\ (R/10\ {\rm km})^2$. This difference comes about because of
extra heating from hot CNO burning during accumulation when hydrogen
is present, leading to earlier ignition, and requiring a lower $\dot
m$ to match the observed recurrence times. For hydrogen to affect the
temperature profile in the layer requires $X\gtrsim 0.03$ and
$Z\gtrsim 3\times 10^{-3}$. If spallation reduces the metallicity in
the accreted layer (Bildsten, Salpeter, \& Wasserman 1992), hydrogen
burning will no longer play a role, in which case the accreting layer
behaves like pure helium.

These conclusions constrain evolutionary models, although both pure
helium evolutionary models and those with hydrogen can be found that
are consistent with the data. A promising way to distinguish between
different evolutionary scenarios is accurate fluence measurements as a
function of $\dot m$. Models with hydrogen predict $\approx 10$\%
variations in burst fluence as recurrence time varies, whereas pure
helium models show a much smaller variation.

For most models, recurrence time is expected to vary as $\approx
1/\dot m$, because the ignition depth (determined by hot CNO burning
or by the flux from the crust) is insensitive to accretion
rate. However, SAS-3 observations showed a much shallower dependence
than the expected $1/\dot m$: Clark et al.~(1977) observed the
persistent flux increase by a factor of $\approx 5$ while $t_{\rm
recur}$ went from $3.4$ to $2.2$ hours. Studies of timing and spectral
behavior of LMXBs indicate that $L_X$ is not necessarily
well-correlated with the underlying accretion rate (e.g.~van der Klis
et al.~1990). The observed behavior then requires $\dot m$ to increase
less quickly than $L_X$. It would be interesting to study burst
properties as a function of an accretion rate indicator (e.g.~position
in the color-color diagram) rather than $L_X$. Variations in heating
with $\dot m$ due to thermal inertia might play some role, something
that should be checked with time-dependent simulations. Alternatively,
these observations might indicate new physics, for example that the
accreting material covers only part of the star at lower accretion
rates, reducing the amount by which the local accretion rate changes
(a possibility suggested for hydrogen accretors by Bildsten 2000). An
interesting exception to the predicted $1/\dot m$ behavior is for
models with $X_0\approx 0.1$, when the time to burn all the hydrogen
is $\approx 3$ hours, in the range of observed recurrence times. For
$X_0\approx 0.1$ and $t_{\rm recur}>3$ hours, the recurrence time is
predicted to increase rapidly with decreasing $\dot m$, because
hydrogen burning no longer occurs throughout the layer. However, this
is even less consistent with the observed variations.

The burst energy is overpredicted in these models, in common with
studies of other bursters. Galloway et al.~(2003) find a factor of
$3$--$4$ discrepancy for GS 1826-24, and following Bildsten (2000)
propose that the accreted fuel only partially covers the neutron star
surface. For 4U~1636-56, Fujimoto et al.~(1987) proposed that vertical
mixing driven by rotational shear in the incoming material led to
early ignition, and reduced burst energies. Here, I find discrepancies
of less than a factor of 2 for 4U~1820-30. This is easily accomodated,
for example, by anisotropic burst emission (Lapidus \& Sunyaev 1985;
Fujimoto 1988), and does not require incomplete spreading of the fuel
or extra mixing. Note however that the discrepancy increases if the
neutron star radius is larger than the $10\ {\rm km}$ assumed here
($E_{\rm burst}\propto R^2$).

I have concentrated on the recurrence time and total burst energy, and
have not discussed other properties, such as the peak luminosity or
burst duration. Type I bursts from 4U~1820-30 generally show radius
expansion, in which the luminosity reaches the Eddington
limit. Kuulkers et al.~(2003) recently surveyed the peak luminosity of
globular cluster bursters. For 4U~1820-30, the average peak flux is
$5.27\pm 0.72\times 10^{-8}\ {\rm erg\ cm^{-2}\ s^{-1}}$, or a
luminosity $3.7\times 10^{38}\ {\rm erg\ s^{-1}}$ for a distance of
$7.6$ kpc. The pure helium Eddington limit is $3.5$--$5\times 10^{38}\
{\rm erg\ s^{-1}}$ for $M=1.4$--$2\ M_\odot$ (not including general
relativistic corrections). When hydrogen is present, the Eddington
luminosity is reduced by a factor $1+X$. However, for small $X$, this
change is much less than the uncertainty in the neutron star mass, and
so does not allow a determination of hydrogen content.

The durations of bursts from 4U~1820-30 are $\approx 20$--$30\ {\rm
s}$ (with exponential decay times $\lesssim 10\ {\rm s}$), consistent
with pure helium burning, in which the helium burns rapidly and the
burst duration is set by the cooling time of the layer. For solar
abundance of hydrogen, slow hydrogen burning via the rp process
(Wallace \& Woosley 1981) extends the decay time to $\sim 100\ {\rm
s}$ (Hanawa, Sugimoto, \& Hashimoto 1983; Wallace \& Woosley 1984;
Hanawa \& Fujimoto 1984; Schatz et al.~1998). However, for
$X_0\lesssim 0.1$, most of the protons are able to burn by direct
capture on carbon. Therefore the small amount of hydrogen considered
in this paper is not expected to substantially increase the burst
duration. This is something that could be addressed by time-dependent
simulations with increasing hydrogen fractions.

\subsection{Superbursts}

In \S 3, I used the $\dot m$'s determined by matching the Type I burst
recurrence times to calculate superburst ignition conditions. For pure
He accretion, the expected superburst recurrence times are $\approx
1$--$2$ years (for $Q_{\rm crust}\approx 0.1$--$0.2$ MeV per nucleon),
much less than found by SB02 (who took $\dot m=\dot m_X$). If hydrogen
is present, the smaller $\dot m$ gives recurrence times $\approx
5$--$10$ years. This approach of simultaneously modelling Type I
bursts and superbursts can be usefully applied to other superburst
sources which exhibit regular Type I bursts. A promising candidate is
KS~1731-260, for which observations of the quiescent flux constrain
$Q_{\rm crust}$ (Rutledge et al.~2002). In addition, these results
emphasize that determining the superburst recurrence time would
strongly constrain the local accretion rate and thermal structure of
the star, and thereby composition of the accreted material.

A direct effect of hydrogen in the accreted material will be to reduce
the amount of carbon made during H/He burning. For pure helium
accretion, SB02 argued that carbon production occurs during stable
burning at high $\dot m$, for which the temperature is low enough to
give a high carbon yield (Brown \& Bildsten 1998). However, if
hydrogen is present, protons will rapidly capture on $^{12}$C,
initiating the hot CNO cycle. In that case, breakout reactions
$^{15}$O$(\alpha$,$\gamma)$ and $^{14}$O$(\alpha$,p$)$ become possible
(Wallace \& Woosley 1981; Schatz et al.~1998), depleting the CNO
abundance, and reducing the carbon yield. For $X_0=0.1$, there are
enough protons for one proton capture on each $^{12}$C. The outcome is
probably different for $X_C$ above and below this value, which is
within the range predicted by models (Podsiadlowski et al.~2002). For
solar abundance of hydrogen, Schatz et al.~(1999) found stable burning
gave $X_C\sim 10$\% (still enough to ignite a superburst, however, as
CB01 showed). Calculations of the carbon production for stable helium
burning with a small amount of hydrogen should be carried out.

Finally, it has been noted that the superburst from 4U~1820-30 is
different from the superburst seen in other sources (Kuulkers et
al.~2002). It was the most energetic, and is the only superburst so
far to reach Eddington luminosity. This is probably due to the low
hydrogen abundance in the accreted material. A lower hydrogen
abundance allows greater carbon production, and makes lighter ashes
(the heavy nuclei are probably iron group with $A\approx 56$ rather
than rp process ashes with $A\sim 100$) reducing the opacity and giving a
deeper ignition. Both these effects give a larger nuclear energy
release, which could explain the more energetic and luminous
superburst.

\subsection{Transition to Stable Burning}

The calculations described here apply only to the low state, when
regular Type I bursts are seen, and do not address the disappearance
of bursts in the high state. This is an important question to
answer. In spherically-symmetric models, burning becomes stable at
high $\dot m$'s because the temperature-sensitivity of the triple
alpha reaction decreases with temperature, and the incoming helium
then burns at the rate at which it accretes (Fujimoto, Hanawa, \&
Miyaji 1981; Ayasli \& Joss 1982; Taam, Woosley, \& Lamb
1996). However, the transition to stable burning for pure helium
accretion occurs at $\dot m_{\rm stab}\approx 2\times 10^6\ {\rm g\
cm^{-2}\ s^{-1}}$ (B95; Bildsten 1998), much higher than 1820-30's
accretion rate. The extra nuclear energy release if hydrogen is
present will decrease $\dot m_{\rm stab}$, but not enough to explain
the observed transition. B95 suggested that the missing piece of
physics was that the bursts become radiative rather than convective in
the high state. The burning then proceeds as a slowly propagating
``fire'' over the stellar surface rather than dramatic Type I
bursts. However, the ignition of a local spot on the surface of the
star is not understood, requiring stabilization of pressure gradients,
perhaps by rapid rotation (Spitkovsky, Levin, \& Ushomirsky 2002;
Zingale et al.~2002).

4U~1820-30 fits into the pattern of bursting seen in other sources: a
transition from frequent, regular bursting to infrequent, irregular
bursting as accretion rate increases (van Paradijs, Penninx, \& Lewin
1988). By comparing {\it BeppoSAX}/WFC data for 9 frequent bursters
(including 4U~1820-30), Cornelisse et al.~(2003) concluded that this
transition occurs at a universal luminosity $L_X\approx 2\times
10^{37}\ {\rm erg\ s^{-1}}$. If this behavior is common to all
bursters, two explanations that have been put forward for hydrogen
accretors may be relevant for 4U~1820-30. First, Narayan \& Heyl
(2003) recently studied the linear stability of quasi-steady burning
shells, finding that some stable burning occurs during accumulation
for accretion rates near the transition to stability. Whether this
result applies to pure helium accretion should be investigated
further. Second, for hydrogen accretors, Bildsten (2000) suggested
that the fraction of the star covered with fresh fuel increases with
increasing global $\dot M$, so that the local accretion rate $\dot m$
{\it decreases}. The observed transition is then from mixed H/He
ignition ($t_{\rm recur}<t_H$) giving frequent, regular bursts to pure
He ignition, giving irregular, less frequent bursts ($t_{\rm
recur}>t_H$). The different composition in 4U~1820-30 makes it
difficult to apply the same explanation, and the regularity of the
bursts perhaps implies complete covering. Increasing area with $\dot
M$ would explain the small change in recurrence time seen by Clark et
al.~(1977). Accurate fluence measurements together with spectral fits
to the radius as a function of $t_{\rm recur}$ are a way to test this
picture.

\acknowledgements I thank Erik Kuulkers, Phillip Podsiadlowski,
Hendrik Schatz and the referee for useful comments, and Lars Bildsten
for stressing that Type I bursts might test evolutionary models for
LMXBs. This work was supported by NASA through Hubble Fellowship grant
HF-01138 awarded by the Space Telescope Science Institute, which is
operated by the Association of Universities for Research in Astronomy,
Inc., for NASA, under contract NAS 5-26555.

\begin{figure}
\plotone{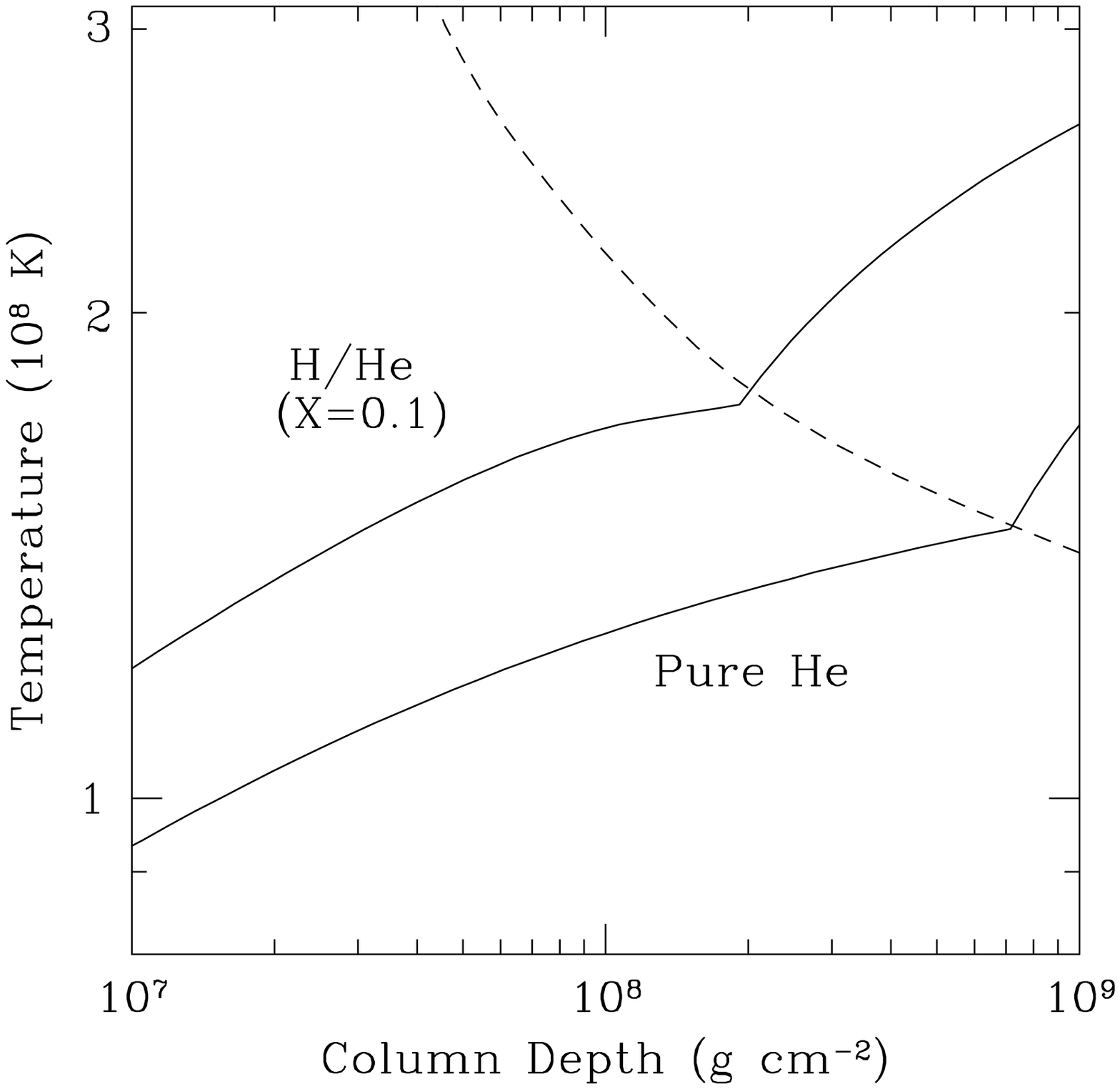}
\caption{Temperature profiles at ignition for accretion of pure He and
H/He with $X=0.1$ at a rate $\dot m=1.2\times 10^4\ {\rm g\ cm^{-2}\
s^{-1}}$. The extra heating from the hot CNO cycle results in a
significantly hotter layer, and earlier ignition. The dashed curve is
the ignition curve for pure He. The model with hydrogen ignites a
little below the ignition curve due to the enhancement of
$\epsilon_{3\alpha}$ by proton captures (see text).\label{fig:prof}}
\end{figure}

\begin{figure}
\plotone{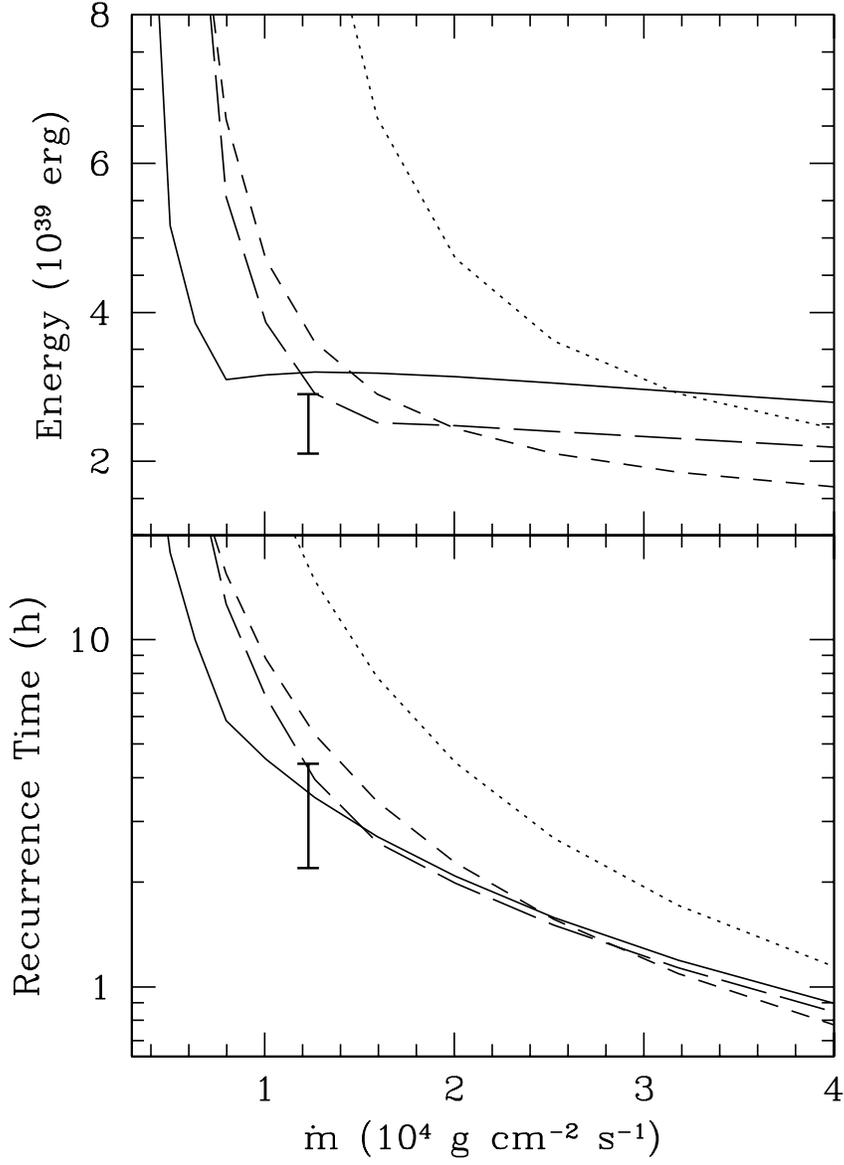}
\caption{ Burst energy and recurrence time for different choices of
instantaneous accretion rate $\dot m$.  I show hydrogen models with
$X=0.1$ (long-dashed line) and $X=0.2$ (solid line), and helium models
with $Q_{\rm crust}=0.1$ MeV per nucleon (dotted line) and $Q_{\rm
crust}=0.2$ MeV per nucleon (short-dashed line). I take
$\dot{\left<m\right>}=2\dot m$ at each $\dot m$, and $Z=0.01$. The
vertical bars show the range of observed properties, and have been
placed at the $\dot m$ inferred from the accretion luminosity $\dot
m_X$.\label{fig:recur}}
\end{figure}

\begin{figure}
\plotone{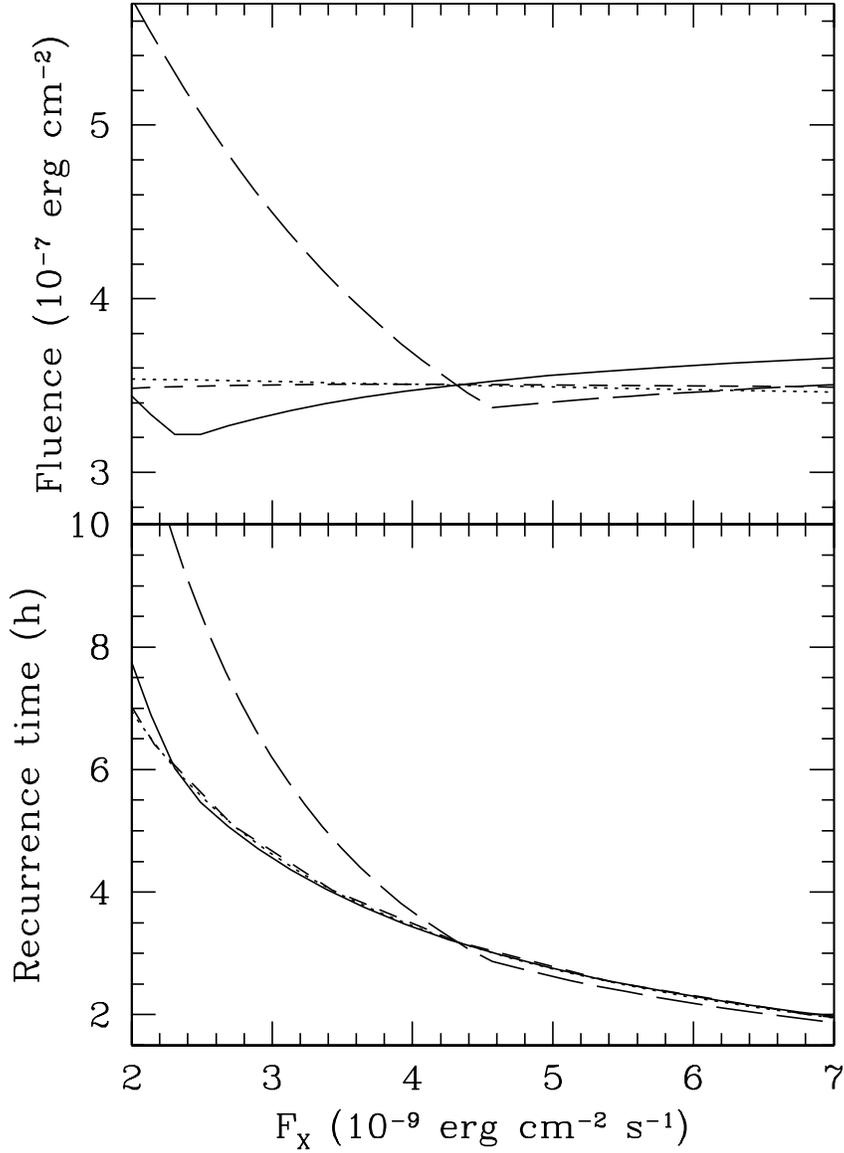}
\caption{Variation of burst properties with persistent flux for a pure
He model (dotted line), a low metallicity hydrogen model (short-dashed
line, $X=0.01$, $Z=10^{-3}$), and two high metallicity hydrogen models
($Z=0.01$; $X=0.1$ long-dashed line; $X=0.2$ solid line). All models
have $Q_{\rm crust}=0.1$ MeV per nucleon. The fluence and recurrence
time have been normalized to the values observed by Haberl et
al.~1987, fluence $3.5\times 10^{-7}\ {\rm erg\ cm^{-2}}$ and $t_{\rm
recur}=3.2$ hours at $F_X=4.3\times 10^{-9}\ {\rm erg\ cm^{-2}\
s^{-1}}$. The kinks in the solid and long-dashed lines occur when the
recurrence time equals the time to burn all the hydrogen (when
$Z=0.01$, this is $\approx 3$ hours for $X=0.1$, $\approx 6$ hours for
$X=0.2$).\label{fig:f4}}
\end{figure}

\begin{figure}
\plotone{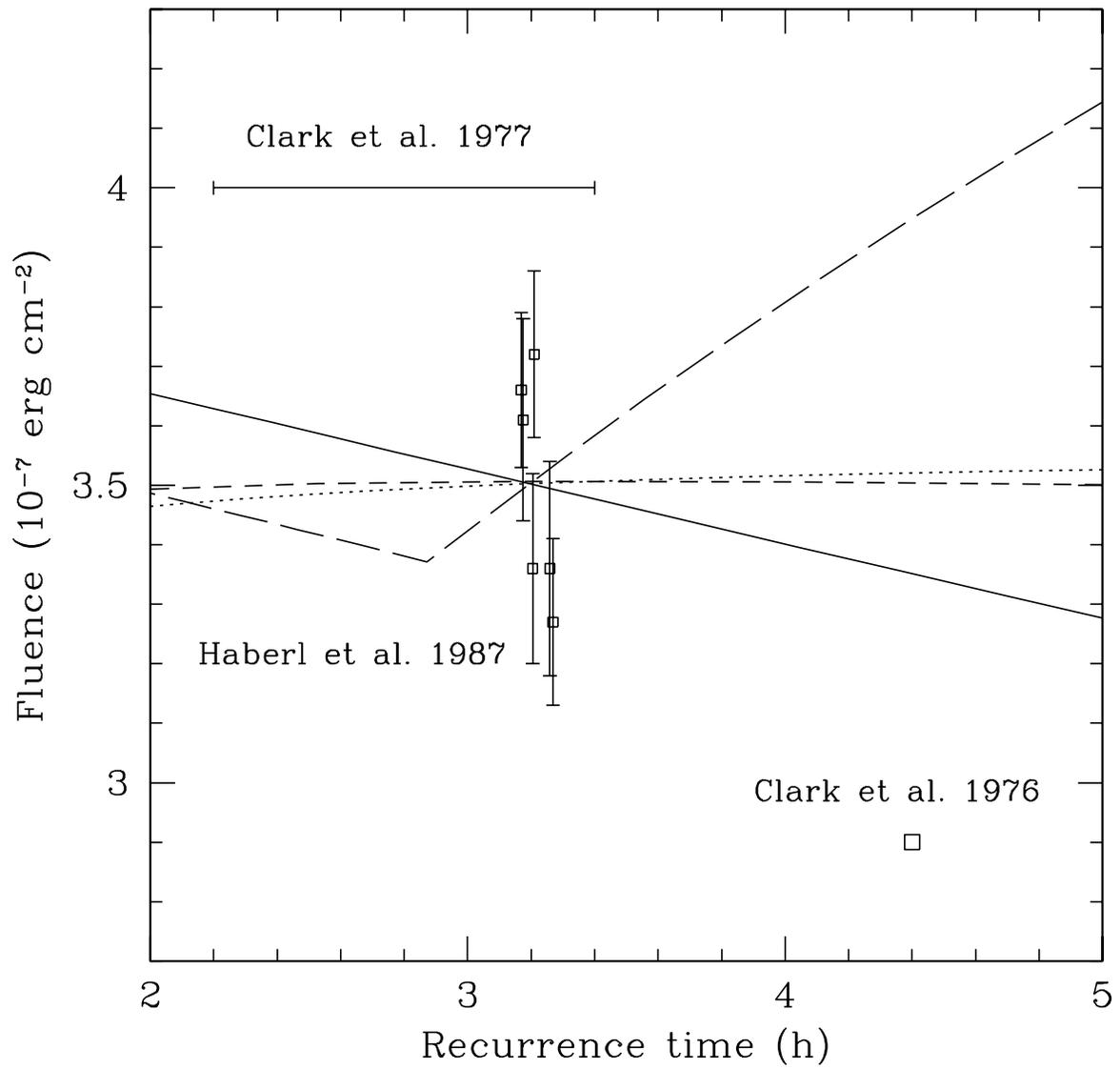}
\caption{ Predicted burst fluence against recurrence time for the
models of Figure \ref{fig:f4}. Little variation of fluence with
$t_{\rm recur}$ is seen for pure He accretion (dotted line), or when
the metallicity in the layer is low ($Z=10^{-3}$, dashed line). The
hydrogen models with $Z=0.01$ show $\approx 10$\% variation in fluence
across the observed range of recurrence times. For $X=0.1$, the slope
changes at $t_{\rm recur}=3$ hours, when the time to burn all the
hydrogen equals the recurrence time.
\label{fig:f5}}
\end{figure}

\clearpage
\begin{deluxetable}{lcccccccccc}
\tablecaption{Ignition conditions for bursts with  $t_{\rm recur}=3.2\ {\rm hours}$\tablenotemark{a}\label{tab:1}}
\tablewidth{0pt}
\tablehead{
\colhead{Model} & \colhead{$X_0$} & \colhead{$Z$} & \colhead{$Q_{\rm crust}$} & \colhead{$\dot m_4$} & \colhead{$y_8$} & \colhead{$T_8$} & \colhead{$X_b$} & \colhead{$\left<X\right>$} & \colhead{$Q_{\rm nuc}$\tablenotemark{b}} & \colhead{$E_{39}$\tablenotemark{c}}
}
\startdata
1 & 0.0 & 0.01 & 0.1 & 2.3 & 2.7 & 1.7 & 0.00 & 0.00 & 1.6 & 3.9\\
2 & 0.0 & 0.01 & 0.2 & 1.7 & 1.9 & 1.8 & 0.00 & 0.00 & 1.6 & 2.8\\
\\
3 & 0.1 & 0.01 & 0.1 & 1.4 & 1.6 & 1.8 & 0.00 & 0.05 & 1.8 & 2.6\\
4 & 0.2 & 0.01 & 0.1 & 1.4 & 1.6 & 1.9 & 0.09 & 0.14 & 2.2 & 3.2\\
5 & 0.3 & 0.01 & 0.1 & 1.4 & 1.6 & 1.9 & 0.19 & 0.24 & 2.6 & 3.9\\
\\
6 & 0.1 & 0.01 & 0.2 & 1.2 & 1.4 & 1.9 & 0.00 & 0.05 & 1.8 & 2.3\\
7 & 0.1 & 0.001 & 0.1 & 2.0 & 2.3 & 1.7 & 0.09 & 0.09 & 2.0 & 4.2\\
\enddata
\tablenotetext{a}{columns are: accreted hydrogen fraction $X_0$; CNO mass fraction $Z$; local accretion rate $\dot m_4=\dot m/10^4\ {\rm g\ cm^{-2}\ s^{-1}}$; ignition column depth $y_8=y/10^8\ {\rm g\ cm^{-2}}$; ignition temperature $T_8=T/10^8\ {\rm K}$; hydrogen mass fraction at the base $X$; mean hydrogen mass fraction $\left<X\right>=\int dy X(y)/y$; nuclear energy release, $Q_{\rm nuc}=1.6+4.0\left<X\right>$ MeV per nucleon; predicted burst energy $E_{39}=E_{\rm burst}/10^{39}\ {\rm erg}$.}
\end{deluxetable}

\begin{deluxetable}{lcccccc}
\tablewidth{0pt}
\tablecaption{Expected properties of Type I X-ray bursts from
evolutionary models\tablenotemark{a}\label{tab:2}}
\tablehead{
\colhead{$\dot{\left<M\right>}_{-9}$} & \colhead{$X_0$} & \colhead{$\dot m_4$} & \colhead{$t_{\rm recur}$ (h)} & \colhead{$X_b$} & \colhead{$Q_{\rm nuc}$} & \colhead{$E_{39}$}
}
\startdata
\multicolumn{7}{c}{Podsiadlowski et al.~(2000)}\\
\hline
8.8 & 0.00 & 2.2 & 3.5 & 0.00 & 1.6 & 4.1\\
4.3 & 0.35 & 1.1 & 4.3 & 0.20 & 2.7 & 4.3\\
11  & 0.18 & 2.8 & 1.4 & 0.13 & 2.2 & 2.9\\
\hline
\multicolumn{7}{c}{Fedorova \& Ergma (1989)}\\
\hline
15  & 0.00 & 3.6 & 1.4 & 0.00 & 1.6 & 2.6\\
4.1 & 0.13 & 1.0 & 4.9 & 0.00 & 1.8 & 3.0\\
5.8 & 0.11 & 1.5 & 2.9 & 0.01 & 1.8 & 2.6\\
7.0 & 0.09 & 1.8 & 2.3 & 0.01 & 1.8 & 2.4\\
9.6 & 0.05 & 2.4 & 1.6 & 0.00 & 1.7 & 2.2\\
\enddata
\tablenotetext{a}{given the time-averaged accretion rate $\dot{\left<M\right>}_{-9}=\dot{\left<M\right>}/10^{-9}\ M_\odot\ {\rm yr^{-1}}$ and hydrogen fraction at the surface of the secondary $X_0$; I assume $\left<\dot m\right>=2\dot m$, giving $\dot m_4=\dot{\left<M\right>}_{-9}/4.0$ for $R=10\ {\rm km}$.}
\end{deluxetable}

\begin{deluxetable}{lcccllclc}
\tablewidth{0pt}
\tablecaption{Superburst Ignition Models\tablenotemark{a}\label{tab:3}}
\tablehead{
\colhead{Model\tablenotemark{b}} & \colhead{$X_0$} & \colhead{$\left<\dot m\right>_4$} & \colhead{$Q_{\rm crust}$} &
\colhead{$t_{\rm recur}$ (yr)} & \colhead{$y_{12}$} & \colhead{$T_8$} &
\colhead{$E_{43}$} & \colhead{$X_{C,{\rm min}}$\tablenotemark{c}}
}
\startdata
SB02 & & 2.4 & 0.1 & 14 & 11 & 3.8 & 3.9 & 0.13\\
\\
1    & 0.0 & 4.6 & 0.1 & 1.9 & 2.8 & 5.0 & 1.0 & 0.08\\
2    & 0.0 & 3.4 & 0.2 & 1.1 & 1.2 & 5.6 & 0.43 & 0.16\\
\\
3    & 0.1 & 2.8 & 0.1 & 9.2 & 8.1 & 4.0 & 3.0 & 0.12\\
6    & 0.1 & 2.4 & 0.2 & 3.7 & 2.8 & 4.9 & 1.0 & 0.19\\
\enddata
\tablenotetext{a}{$\left<\dot m\right>_4=\left<\dot m\right>/10^4\ {\rm g\ cm^{-2}\ s^{-1}}$; $y_{12}=y_{\rm ign}/10^{12}\ {\rm g\ cm^{-2}}$; $E_{43}=E_{\rm superburst}/10^{43}\ {\rm erg}$. We assume a composition of 30\% $^{12}$C, 70\% $^{56}$Fe, and calculate the total energy release $E_{\rm superburst}=4\pi R^2y_{\rm ign}Q_{\rm nuc}$ assuming $Q_{\rm nuc}=1$ MeV per nucleon. }
\tablenotetext{b}{The model number refers to Table \ref{tab:1}, except for SB02 which refers to Strohmayer \& Brown 2002. Models 1 and 2 are pure He accretion; models 3 and 6 have $X=0.1$.}
\tablenotetext{c}{The minimum mass fraction of carbon required for unstable ignition. The carbon burns stably during accumulation for lower $X_C$.}
\end{deluxetable}

\end{document}